\documentclass[preprint,showpacs,preprintnumbers,amsmath,amssymb,nofootinbib]{revtex4}

\usepackage{epsfig}
\usepackage{graphicx}% Include figure files
\usepackage{dcolumn}% Align table columns on decimal point
\usepackage{bm}% bold math

\def\DESepsf(#1 width #2){\epsfxsize=#2 \epsfbox{#1}}
\newcommand{\im}{{\rm Im}}
\newcommand{\out}{{\rm out}}
\begin{document}

\preprint{April, 2005}

\title{Implications of $\overline B\to D^0 h^0$ Decays on
$\overline B\to D\overline K,~\overline D\, \overline K$ Decays}
%
% Force line breaks with \\

\author{Chun-Khiang Chua}
\affiliation{Institute of Physics, Academia Sinica, Taipei, Taiwan
115, Republic of China}
\author{Wei-Shu Hou}%
\affiliation{%
Physics Department, National Taiwan University, Taipei, Taiwan
10764, Republic of China
}%


\begin{abstract}
The recently observed color suppressed $\overline B{}^0 \to
D^0\pi^0$, $D^0\eta^{(\prime)}$, $D_s^+ K^-$ and $D^0\overline K
{}^0$ decay modes all have rates larger than expected, hinting at
the presence of final state interactions. We study rescattering
effects in $\overline B\to DP$, $D\overline K$ and $\overline D\,
\overline K$ modes in the quasi-elastic approach, which is
extended to accommodate $D^0\eta'$ without using U(3) symmetry.
The $\overline D {}^0 \overline K$ modes are of interest in the
determination of the unitarity angle $\phi_3/\gamma$. The updated
$DP$ data are used to extract the effective Wilson coefficients
$a^{\rm eff}_1\simeq 0.92$, $a^{\rm eff}_2\simeq 0.22$, three
strong phases $\delta\simeq 62^\circ$, $\theta\simeq 24^\circ$,
$\sigma\simeq 127^\circ$, and the mixing angle $\tau\simeq
2^\circ$. The values of $\delta$ and $\theta$ are close to our
previous results. The smallness of $\tau$ implies small mixing of
$D^0\eta_1$ with other modes. Predictions for $D^0K^-$, $D^+ K^-$
and $D^0\overline K {}^0$ agree with data. The framework applies
to $\overline B \to \overline D\, \overline K$, and rates for
$\overline D {}^0 K^-$, $D^- K^0$, $D_s^-\pi^0$, $D_s^-\eta$ and
$D_s^-\eta'$ modes are predicted. From $B^-\to \overline D{}^0K^-$
and $D^0K^-$ rates, we find $r_B=0.09\pm0.02$.
\end{abstract}

  %Bottom mesons
%\pacs{ %Valid PACS appear here
\maketitle

\section{Introduction}

The color-suppressed decays $\overline B^0\to
D^{(*)0}\pi^0$~\cite{BelleBDpi,CLEOBDpi} and
$D^0\eta,D^0\omega$~\cite{BelleBDpi} were observed for the first
time in 2001. Recently, improved measurements of $\overline B^0\to
D^{(*)0}(\pi^0,\eta,\omega)$~\cite{BaBarBDpi,Belleupdate} and the
first observation of the $D^0\eta'$ mode have been reported by
BaBar~\cite{BaBarBDpi} and confirmed by
Belle~\cite{Schumann:2005ej}. Other color suppressed modes, such
as $D_s K^-$ and $D^0 \overline K{}^0$, have also been
observed~\cite{DsK,D0K0}. All these modes
have branching ratios %with central
%values between $1.7\times 10^{-4}$ and $4.2\times 10^{-4}$. The fact that they
that are significantly larger than earlier theoretical
expectations based on naive factorization, indicating the presence
of non-vanishing strong phases, which has attracted much
attention~\cite{Xing:2001nj,Cheng:2001sc,Neubert:2001sj,Chua:2001br,Chiang:2002tv,SCET,pQCD,
Gronau:2002mu,Fleischer:2003,Wolfenstein:2003pc,Cheng:2004ru}.
Shortly after the first observation of the color suppressed modes
became known, we proposed~\cite{Chua:2001br} a quasi-elastic final
state rescattering (FSI) picture, where the enhancement of color
suppressed $D^0 h^0$ modes can be understood as rescattering from
the color allowed $D^+\pi^-$ final state. The framework is
applicable to
 $\overline B\to D\overline K$, $\overline
D\,\overline K$ decays.

%%%%%%%%%%%%%%%%%%%%%%%%%%%%%%%%%%%%%%%%%%%%%%%%%%%%%
%Fig 1%
%%%%%%%%
\begin{figure}[t!]
\centerline{\hskip-1.1cm \DESepsf(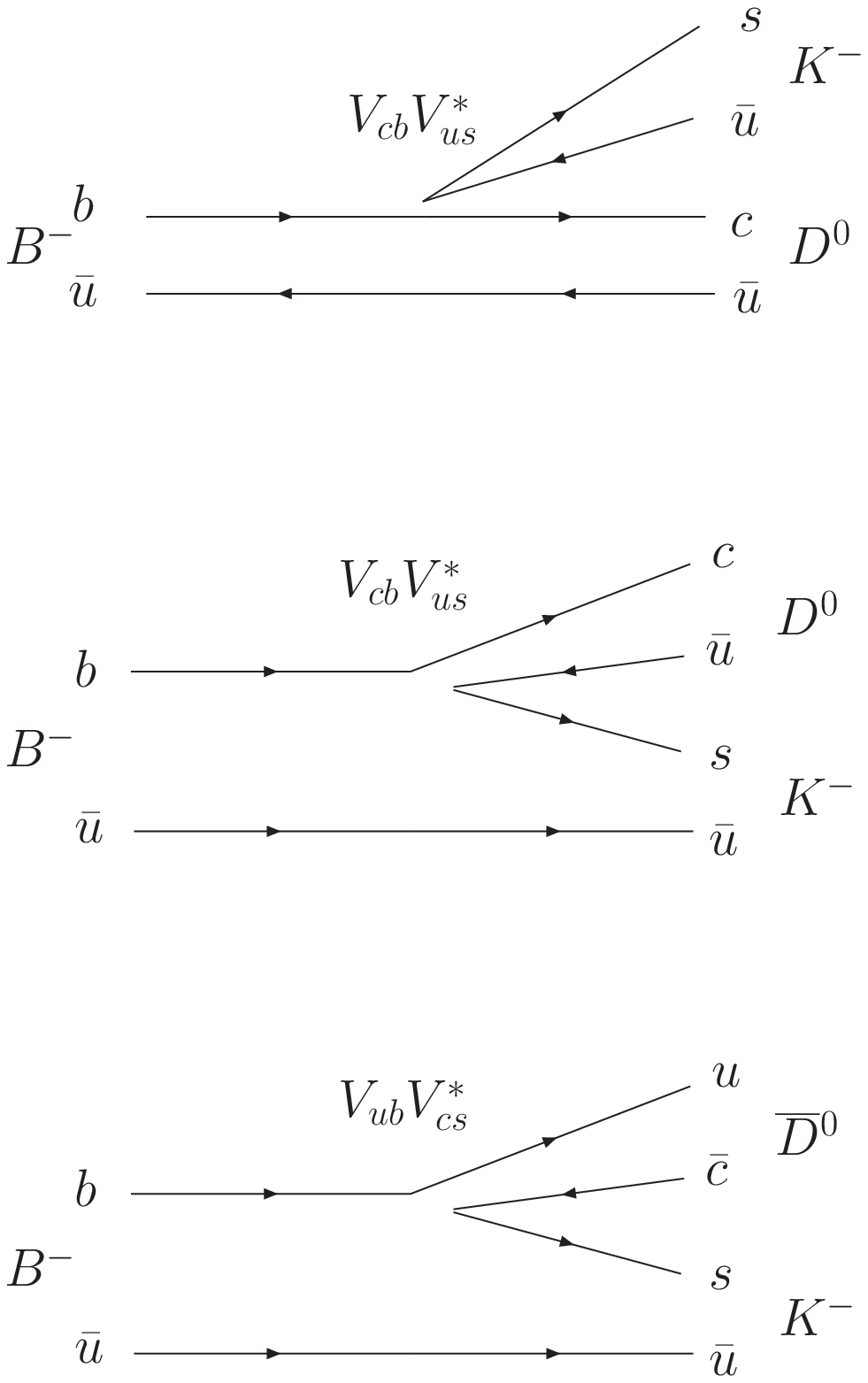 width 9cm)}
\caption{Color-allowed and color-suppressed $B^-\to D^0 K^-$
decay, and color and CKM-suppressed $B^-\to\overline D {}^0 K^-$
decay.
 }
 \label{fig:DKDbarK}
\end{figure}
%%%%%%%%%%%%%%%%%%%%%%%%%%%%%%%%%%%%%%%%%%%%%%%%%%%%%

The color-allowed $B^-\to D^0K^-$ and color-suppressed $\overline
D {}^0 K^-$ decays are of interest for the determination of the
unitary phase angle $\phi_3(\gamma)\equiv \arg V_{ub}^*$, where
$V$ is the Cabibbo-Kobayashi-Maskawa (CKM) quark mixing matrix. To
be more specific, the amplitude ratio $r_B$ and the strong phase
difference $\delta_B$ for $\overline D {}^0 K^-$ and $D^0 K^-$
decay modes, which are governed by different CKM matrices as
depicted in Fig.~\ref{fig:DKDbarK}, are defined as
 \begin{equation}
 r_B=\left|\frac{A(B^-\to \overline D {}^0 K^-)}{A(B^-\to D {}^0
 K^-)}\right|,
 \qquad
 \delta_B=\arg\left[\frac{e^{i\phi_3}
  A(B^-\to \overline D {}^0 K^-)}{A(B^-\to D {}^0 K^-)}\right].
 \label{eq:rB}
 \end{equation}
The weak phase $\phi_3$ is removed from $A(B^-\to \overline D {}^0
K^-)$ in defining $\delta_B$. The $r_B$ and $\delta_B$ parameters
are common to the $\phi_3$ determination methods of
Gronau-London-Wyler (GLW)~\cite{GLW}, Atwood-Dunietz-Soni
(ADS)~\cite{ADS} and ``$DK$ Dalitz
plot"~\cite{Dalitz,DalitzBelle0304}, where one exploits the
interference effects of $B^-\to D^0 K^-\to f_{\rm CP} K^-$ and
$B^-\to \overline D {}^0 K^-\to f_{\rm CP}K^-$ amplitudes. Note
that the $r_B$ parameter, which governs the strength of
interference, is both color and CKM suppressed, hence hard to
measure directly.

Through the $DK$ Dalitz plot method, the BaBar and Belle
experiments already find $\gamma=70^\circ\pm
44^\circ\pm10^\circ\pm10^\circ$ and $\phi_3=64^\circ\pm
19^\circ\pm13^\circ\pm11^\circ$, respectively~\cite{HFAG,gammaDK},
where the last error comes from modelling of $D$ decay resonances
across the Dalitz plot for, e.g. $D^0 \to K_S\pi^+\pi^-$. Although
similar results on $\phi_3$ are obtained, the corresponding $r_B$
values are quite different for BaBar and Belle. Belle reports
$r_B=0.21\pm0.08\pm0.03\pm0.04$, while BaBar gives $r_B<0.19$ at
90\% confidence level. Note that an average $r_B=0.10\pm 0.04$ is
found by the UT$_{fit}$ group, by combining analyses using all
three methods~\cite{UTfit}. As the strength of interference is
governed by the size of $r_B$, the larger error in the $\gamma$
value of BaBar reflects the smallness of their $r_B$. Given the
present experimental situation that Belle and BaBar have quite
different $r_B$ values and the critical role it takes in
$\phi_3/\gamma$ extraction, it is important to give a theoretical
or phenomenological prediction of $r_B$ and $\delta_B$. Not much
work has so far been done.~\footnote{In the preparation of this
paper, we note that a calculation in the pQCD approach has been
reported~\cite{Keum}.}

The enhancement in rates of the color-suppressed $DP$ modes could
imply~\cite{Gronau:2002mu} a larger $r_B$. It is thus of interest
to study $DP$ and $\overline D\,\overline K$ modes together. In
fact, it was noted in Ref.~\cite{Chua:2001br} that the
quasi-elastic approach used in the $DP$ system can be applied to
the $\overline D\, \overline K$ system. Rescattering parameters
are basically non-perturbative and can only be fitted from data.
We shall use the $\overline B \to D\pi,\,D\eta^{(\prime)},\,D_s
\overline K$ decay rates to extract the rescattering parameters,
which are
then used to predict $D\overline K$ %(subset of $DP$)
and $\overline D\, \overline K$ rates. In this way, we are able to
deduce what value of $r_B$ is preferred in the quasi-elastic
rescattering scenario.

Our previous analysis on FSI effects in $D^0h^0$ modes was based
on early data. Recent experimental updates show some variations.
For example, the $D^0\pi^0$ rate has dropped while the $D^0\eta$
rate is larger. Furthermore, the $D^0\eta'$ mode has finally been
measured. In our earlier study~\cite{Chua:2001br}, because of the
absence of the $D^0\eta'$ mode, we ignored it and approximated
$D^0\eta$ by $D^0\eta_8$, by argument of the U$_A(1)$ anomaly and
small singlet--octet (or $\eta$--$\eta^\prime$) mixing. The same
approach was applied to the study of the charmless
case~\cite{Chua:2002wk}. Given the long standing problem of the
$B\to \eta'K$ rate, it is of interest to clarify the $\eta_1$
issue in $B$ decays~\cite{Keta}. With the emerging new data, it is
time to update the quasi-elastic rescattering approach, and verify
the approximations made.

In Sec. II we extend the quasi-elastic rescattering framework to
include $D^0\eta'$. Because of the U$_A(1)$ anomaly, we use SU(3)
rather than U(3) symmetry. As a consequence, we need three phases
and one mixing angle as rescattering parameters. These same
parameters also enter the rescattering in the $D\overline K$ and
$\overline D\,\overline K$ systems. In Sec.~III we carry out a
numerical study. The effective Wilson coefficients and
rescattering parameters are obtained by using current $\overline
B\to DP$ data. The $\overline B\to D\overline K$ rates are then
predicted and compared with current data. We then proceed to study
the $\overline B\to \overline D\,\overline K$ system and make
predictions for $r_B$ and $\delta_B$. The conclusion is then
offered in Sec. IV.

\section{Final State Rescattering Framework}

\subsection{Quasi-elastic Rescattering}

Let $H_{\rm W}$ be the weak decay Hamiltonian. In the absence of
weak phase (or if they are factored out),  $H_{\rm W}$ is
time-reversal invariant. By using time reversal invariance of
$H_{\rm W}$ and the optical theorem, we have (see, for example,
\cite{Suzuki:1999uc,Chua:2001br})
\begin{equation}
 2\,\im\, \langle i;\out|H_{\rm W}|B\rangle
  = \sum_j {\cal T}^*_{ji}\, \langle j;\out|H_{\rm W}|B\rangle,
 \label{eq:ImA}
\end{equation}
where ${\cal T}$ is the ${\cal T}$-matrix of strong scattering,
and the phase convention $T\,|{\rm in}\rangle=|\out\rangle$ under
time-reversal operation is used. This is the master formula of FSI
in $B$ decay. In particular, for $B$ decay to two body final state
with momentum ($p_1,p_2$), we have, %as we illustrate in Fig. 2.
\begin{eqnarray}
-2i\,{\rm Im}\,A(p_B\to p_1 p_2) &=& \sum_j\left(\prod_{k=1}^j
\int {d^3q_k\over (2\pi)^3 2 E_k}\right)
             (2\pi)^4 \delta^4 \left(p_1+p_2-\sum^j_{k=1} q_k\right)
\nonumber \\
&&\qquad\qquad \times M^*(p_1 p_2\to \{q_k\}) A(p_B\to \{q_k\}) ,
\label{eq:optical}%{eq:optical}
\end{eqnarray}
where the optical theorem is used to all orders of the strong
interaction, but only to first order of the weak interaction.
%Note that $A(p_B\to p_1 p_2)$ and $A(p_B\to \{q_k\})$ are governed
%by the same effective Hamiltonian as explicitly shown in
%Eq.~(\ref{eq:ImA}).
Eq.~(\ref{eq:optical}) relates the imaginary part of the two body
decay amplitude to the sum over all possible $B$ decay final
states $\{q_k\}$, followed by $\{q_k\}\to p_1p_2$ rescattering.
The solution to the above equation is
\begin{equation}
A={\cal S}^{1/2} A^0, \label{eq:A=S12A0}
\end{equation}
where $A^0$ is real, and ${\cal S}=1+i\,{\cal T}$. The weak decay
picks up strong scattering phases~\cite{Watson:1952ji}.

It has been pointed out that \cite{Donoghue:1996hz} elastic
rescattering effects may not be greatly suppressed at $m_B$ scale,
while inelastic rescattering contributions may be important. But
we would clearly lose control if the full structure shown in
Eq.~(\ref{eq:optical}) is employed. Even if all possible $B$ decay
rates can be measured, it would be impossible to know the phases
of each amplitude. Furthermore, we know very little about the
strong rescattering amplitudes. However, in Eq.~(\ref{eq:optical})
the subset of two body final states that may be reached via {\it
elastic} rescatterings stand out compared to inelastic channels.
It has been shown from duality arguments \cite{Gerard:1991ni}, as
well as a statistical approach \cite{Suzuki:1999uc}, that
inelastic FSI amplitudes tend to cancel each other and lead to
small FSI phases. We shall therefore separate $\{q_k\}$ into two
body elastic channels plus the rest, and concentrate on the
contribution of the former.

We will consider $DP$ final states, where $P$ stands for a
pseudoscalar meson. We consider $D$ and $P$ within SU(3)
multiplets, for example the $D$ anti-triplet of $D^0$, $D^+$ and
$D_s^+$, the $\Pi$ octet that contains pions, kaons and the $\eta$
($\eta_8$ component) meson, as well as the $\eta_1$ (mixing of
physical $\eta$ and $\eta^\prime$). Thus, we call this
quasi-elastic rescattering.

\subsection{Rescattering Formalism including $D^0\eta_1$}

In Ref.~\cite{Chua:2001br} we treated $D\Pi \to D\Pi$ rescattering
and considered only the $\eta_8$, since at that time $\overline
B{}^0\to D^0\eta^\prime$ was not yet reported. Here we wish to
extend the formalism to include $D^0\eta_1$, so we can treat the
physical $D^0\eta$ and $D^0\eta^\prime$ final states. The
quasi-elastic strong rescattering for $C = +1$, $S = 0$ final
states can be written as
\begin{equation}
\left(
\begin{array}{l}
A_{D^+\pi^-}\\
A_{D^0\pi^0}\\
A_{D^+_s K^-}\\
A_{D^0\eta_8}\\
A_{D^0\eta_1}
\end{array}
\right) ={\cal S}^{1/2}\, \left(
\begin{array}{l}
A^0_{D^+\pi^-}\\
A^0_{D^0\pi^0}\\
A^0_{D^+_s K^-}\\
A^0_{D^0\eta_8}\\
A^0_{D^0\eta_1}
\end{array}
\right),
 \label{eq:FSIDpi}
\end{equation}
where ${\cal S}^{1/2}=(1+i{\cal T})^{1/2}=1+i{\cal T}^\prime$,
with
\begin{equation}
{\cal T}=\left(
\begin{array}{ccccc}
r_0+r_a
       &\frac{r_a-r_e}{\sqrt2}
       &r_a
       &\frac{r_a+r_e}{\sqrt6}
       &\frac{\bar r_a+\bar r_e}{\sqrt3}
       \\
\frac{r_a-r_e}{\sqrt2}
       &r_0+\frac{r_a+r_e}{2}
       &\frac{r_a}{\sqrt2}
       &\frac{r_a+r_e}{2\sqrt3}
       &\frac{\bar r_a+\bar r_e}{\sqrt6}
       \\
r_a
       &\frac{r_a}{\sqrt2}
       &r_0+r_a
       &\frac{r_a-2 r_e}{\sqrt6}
       &\frac{\bar r_a+\bar r_e}{\sqrt3}
       \\
\frac{r_a+r_e}{\sqrt6}
       &\frac{r_a+r_e}{2\sqrt3}
       &\frac{r_a-2 r_e}{\sqrt6}
       &r_0+\frac{r_a+r_e}{6}
       &\frac{\bar r_a+\bar r_e}{3\sqrt2}
       \\
\frac{\bar r_a+\bar r_e}{\sqrt3}
       &\frac{\bar r_a+\bar r_e}{\sqrt6}
       &\frac{\bar r_a+\bar r_e}{\sqrt3}
       &\frac{\bar r_a+\bar r_e}{3\sqrt2}
       &\tilde r_0+\frac{\tilde r_a+\tilde r_e}{3}
\end{array}
\right), \label{eq:TDpi}
\end{equation}
and ${\cal T}^\prime$ has the same structure as ${\cal T}$ by
SU(3) symmetry, but with $r_i$ replaced by $r_i^\prime$. In
addition, we have $A_{D^0\pi^-}=(1+i r'_0+ir'_e) A^0_{D^0\pi^-}$.
The $r_i^{(\prime)}$s are discussed below.

%%%%%%%%%%%%%%%%%%%%%%%%%%%%%%%%%%%%%%%%%%%%%%%%%%%%%
%Fig 2%
%%%%%%%%
\begin{figure}[t!]
\centerline{\hskip-1.1cm \DESepsf(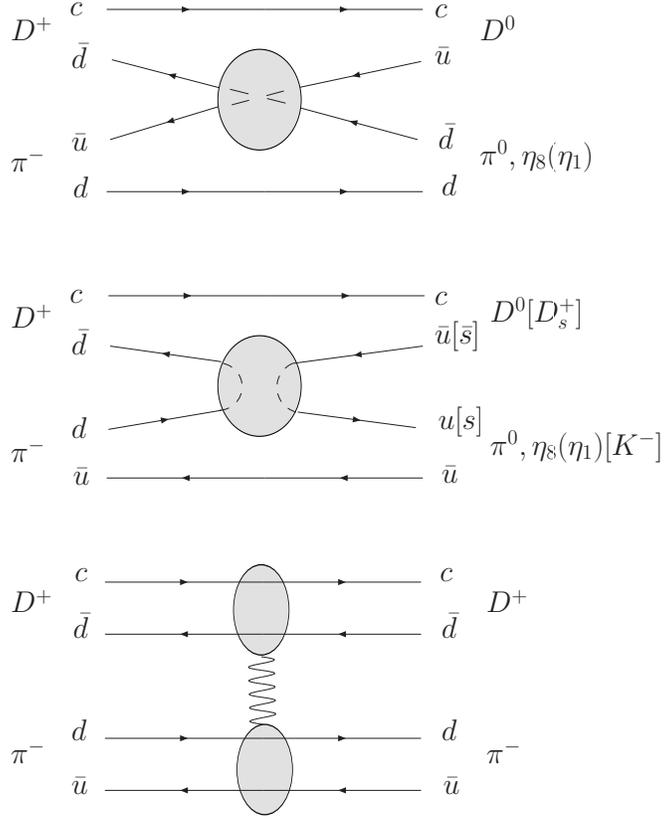 width 10cm)}
\caption{Pictorial representation (from top to bottom) of
   charge exchange $r_e$ ($\bar r_e$),
   annihilation $r_a$ ($\bar r_a$) and
   singlet exchange $r_0$ ($\bar r_0$)
   for $DP$ (re)scattering.
} \label{fig:r2r1r0}
\end{figure}
%%%%%%%%%%%%%%%%%%%%%%%%%%%%%%%%%%%%%%%%%%%%%%%%%%%%%

Eq.~(\ref{eq:TDpi}) is obtained by separating the $D\Pi\to D\Pi$
%(where $\Pi$ is the pseudoscalar octet)
scattering amplitude $M$ into three independent components,
$M_{0,a,e}$, and defining
\begin{equation}
r_{i}\equiv \int {d^3q_1\over (2\pi)^3 2 E_1}{d^3q_2\over (2\pi)^3
2 E_2} \; (2\pi)^4\delta^4 (p_1+p_2-q_1-q_2)\,M_{i}(p_1p_2\to
q_1q_2). \label{eq:ridef}
\end{equation}
Note that the $M_0$, $M_a$ and $M_e$ amplitudes correspond
respectively to the three independent SU(3) combinations ${\rm
Tr}(D_{\rm in} D_{\rm out}) {\rm Tr}(\Pi_{\rm in}\Pi_{\rm out})$,
${\rm Tr}(D_{\rm in} \Pi_{\rm in}\Pi_{\rm out} D_{\rm out})$ and
${\rm Tr}(D_{\rm in} \Pi_{\rm out}\Pi_{\rm in} D_{\rm out})$ of
$D_{\rm in}\Pi_{\rm in}\to D_{\rm out}\Pi_{\rm out}$ scattering.
For $D\Pi\leftrightarrow D^0\eta_1$ and $D^0\eta_1\leftrightarrow
D^0\eta_1$ scattering, we denote the corresponding integrals as
$\bar r_i$ and $\tilde r_i$, respectively. Had U(3) rather than
SU(3) symmetry held in the light pseudoscalar sector, $\bar r_i$
and $\tilde r_i$ would have been identified with $r_i$.

We give a pictorial representation of $r_e$, $r_a$, $r_0$ in
Fig.~\ref{fig:r2r1r0}, which can be seen as corresponding to
charge exchange, annihilation, and flavor singlet exchange,
respectively. The coefficients $r_i$ in Eq.~(\ref{eq:TDpi}) can be
reproduced easily using this pictorial approach by matching the
flavor wave function coefficients. For example, we have
$(r_a-r_e)/\sqrt2$ for $D^+\pi^- \to D^0\pi^0$ rescattering.
Exchange rescattering ($r_e$), the first diagram of
Fig.~\ref{fig:r2r1r0}, projects out the $d\bar d$ component of
$\pi^0$ on the right hand side. This give a $-1/\sqrt2$ factor
from the $\pi^0$ wave function. Similarly, the second diagram of
Fig.~\ref{fig:r2r1r0} projects out the $u\bar u$ component of
$\pi^0$, hence gives $r_a/\sqrt2$. These diagrams also provide
further information. For example, it is easy to see from the
second diagram that annihilation rescattering ($r_a$) is
responsible for $D^+\pi^-\to D^+_sK^-$, since there is no $s$
quark before rescattering. The $M_i$s in Eq.~(\ref{eq:ridef}) can
be understood analogously. For example, the $M_e$ operator
$(D_{\rm in})_i (\Pi_{\rm out})^i_k(\Pi_{\rm in})^k_j (D_{\rm
out})^j$ corresponds to~\footnote { Superscripts and subscripts
are assigned according to the field convention with $q^i$ and
$\bar q_j$ as quark and antiquark fields, respectively.
 }
the $D(c\bar q_i)\Pi(\bar q_j q^k)\to D(c\bar q^j)\Pi(\bar q^i
q_k)$ process with the exchange of $i$th and $j$th antiquark,
hence it is the exchange rescattering operator.

%%%%%%%%%%%%%%%%%%%%%%%%%%%%%%%%%%%%%%%%%%%%%%%%%%%%%
%Fig 3%
%%%%%%%%
\begin{figure}[t!]
\centerline{\hskip-1.1cm \DESepsf(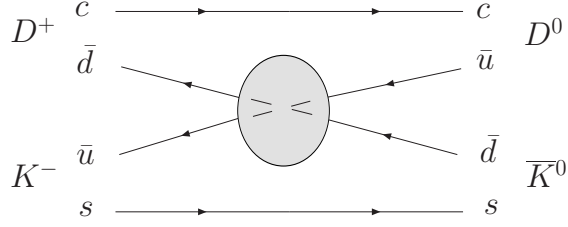 width 8cm)}
\caption{Pictorial representation  of
   charge exchange $r_e$ in $D^+ K^-\to D^0 \overline K {}^0$ (re)scattering.
} \label{fig:reDK}
\end{figure}
%%%%%%%%%%%%%%%%%%%%%%%%%%%%%%%%%%%%%%%%%%%%%%%%%%%%%

For Cabibbo suppressed $\overline B \to D^{+}K^-$ and
$D^{0}\overline K^0$ modes, we have
\begin{eqnarray}
\left(
\begin{array}{l}
A_{D^+K^-}\\
A_{D^0\overline K{}^0}
\end{array}
\right) &=& \left(
\begin{array}{cc}
1+i r^\prime_0 &i r^\prime_e\\
i r^\prime_e &1+i r^\prime_0
\end{array}
\right) \left(
\begin{array}{l}
A^0_{D^+K^-}\\
A^0_{D^0\overline K{}^0}
\end{array}
\right) , \label{eq:A=S12ADK}
\end{eqnarray}
which can be easily obtained by using the pictorial approach shown
in Fig.~\ref{fig:reDK}. It is clear that ``annihilation"
rescattering is impossible in this case. For $D^0 K^-$, we have
$A_{D^0 K^-}=(1+i r'_0+ir'_e) A^0_{D^0 K^-}$.

\begin{figure}[t!]
\centerline{\hskip-1.1cm \DESepsf(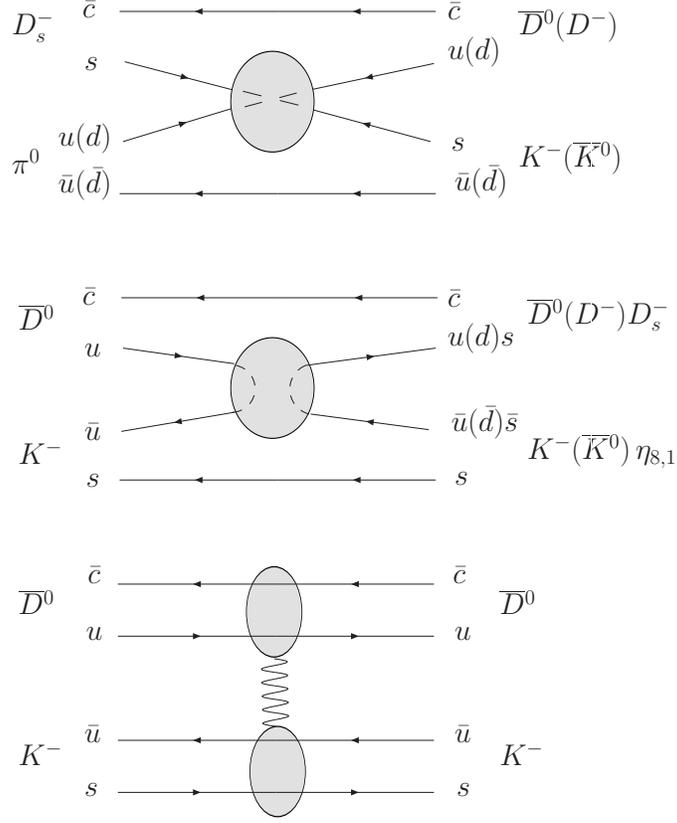 width 10cm)}
\caption{Pictorial representation of
   charge exchange,
   annihilation and
   singlet exchange
   (re)scatterings involving the $\overline D {}^0 K^-$ state.
} \label{fig:rerar0DK}
\end{figure}

For the $B^-\to \overline D\,\overline K$ decays, we have
\begin{equation}
\left(
\begin{array}{l}
A_{\overline D {}^0 K^-}\\
A_{D^- \overline K{}^0}\\
A_{D^-_s \pi^0}\\
A_{D^-_s\eta_8}\\
A_{D^-_s\eta_1}
\end{array}
\right) ={\cal S}^{1/2}\, \left(
\begin{array}{l}
A^0_{\overline D {}^0 K^-}\\
A^0_{D^- \overline K{}^0}\\
A^0_{D^-_s \pi^0}\\
A^0_{D^-_s\eta_8}\\
A^0_{D^-_s\eta_1}
\end{array}
\right),
 \label{eq:FSIDbarK}
\end{equation}
where ${\cal S}^{1/2}=(1+i{\cal T})^{1/2}=1+i{\cal T}^\prime$,
with
\begin{equation}
{\cal T}=\left(
\begin{array}{ccccc}
r_0+r_a
       &r_a
       &\frac{r_e}{\sqrt2}
       &\frac{r_e-2 r_a}{\sqrt6}
       &\frac{\bar r_e+\bar r_a}{\sqrt3}
       \\
r_a
       &r_0+r_a
       &-\frac{r_e}{\sqrt2}
       &\frac{r_e-2 r_a}{\sqrt6}
       &\frac{\bar r_a+\bar r_e}{\sqrt3}
       \\
\frac{r_e}{\sqrt2}
       &-\frac{r_e}{\sqrt2}
       &r_0
       &0
       &0
       \\
\frac{r_e-2 r_a}{\sqrt6}
       &\frac{r_e-2 r_a}{\sqrt6}
       &0
       &r_0+\frac{2}{3}(r_a+r_e)
       &-\frac{\sqrt2}{3}(\bar r_a+\bar r_e)
       \\
\frac{\bar r_e+\bar r_a}{\sqrt3}
       &\frac{\bar r_a+\bar r_e}{\sqrt3}
       &0
       &-\frac{\sqrt2}{3}(\bar r_a+\bar r_e)
       &\tilde r_0+\frac{\tilde r_a+\tilde r_e}{3}
\end{array}
\right), \label{eq:TDbarK}
\end{equation}
as one can easily verify using the pictorial approach shown in
Fig.~\ref{fig:rerar0DK}. The zeros are a consequence of assuming
isospin symmetry.
It is important to note that, due to charge conjugation invariance
and SU(3) symmetry of the strong interactions, these
$r^{(\prime)}_i$, $\bar r^{(\prime)}_i$ and $\tilde
r^{(\prime)}_i$ coefficients are identical to those in
Eqs.~(\ref{eq:TDpi}) and (\ref{eq:A=S12ADK}).

%%%%%%%%%%%%%%%%%%%%%%%%%%%%%%%%%%%%%%%%%%%%%%%%%%%%%
%Fig 4%
%%%%%%%%
%%%%%%%%%%%%%%%%%%%%%%%%%%%%%%%%%%%%%%%%%%%%%%%%%%%%%

By solving ${\cal S}^\dagger {\cal S}=1$, we
obtain~\cite{Chua:2001br}

 \begin{eqnarray}
 (1+i r_0)&=&\frac{1}{2}(1+e^{2i\delta}),
 \nonumber\\
 i r_e &=&\frac{1}{2}(1-e^{2i\delta}),
 \nonumber\\
 i r_a &=&\frac{1}{8}(3 {\cal U}_{11}-2 e^{2i\delta}-1),
 \nonumber\\
 i(\bar r_a+\bar r_e)&=&\frac{3}{2\sqrt2} {\cal U}_{12},
 \nonumber\\
 i(\tilde r_0+\frac{\tilde r_a+\tilde r_e}{3})&=&{\cal U}_{22}-1,
 \label{eq:solution}
 \end{eqnarray}
where
\begin{equation}
{\cal U}={\cal U}^T=\left(
\begin{array}{cc}
\cos\tau
       &\sin\tau
       \\
-\sin\tau
       &\cos\tau
\end{array}
\right)
\left(
\begin{array}{cc}
e^{2i\theta}
       &0
       \\
0
       &e^{2i\sigma}
\end{array}
\right)
\left(
\begin{array}{cc}
\cos\tau
       &-\sin\tau
       \\
\sin\tau
       &\cos\tau
\end{array}
\right), \label{eq:U}
\end{equation}
where we have set the overall phase factor ($1+ir_0+ir_e$) in
${\cal S}$ to unity. This phase convention is equivalent to
choosing the $A_{D^0\pi^-}$ amplitude to be real.

We stress that the above solution satisfies ${\cal S}^\dagger
{\cal S}=1$ in all three cases of $DP$ ($P$ now stands for
strangeness 0 pseudoscalar), $D\overline K$ and $\overline
D\,\overline K$. The $r'_i$, $\bar r'_i$ and $\tilde r'_i$ in
${\cal S}^{1/2}$ can be obtained by using the above formulas with
phases ($\delta,\,\theta,\,\sigma$) reduced by half. We need three
phases and one mixing angle to specify FSI effects since one does
not have nonet symmetry in the light pseudoscalar sector. An extra
phase as well as mixing angle arise from including $\eta_1$ in our
analysis. We will resort to data to see how far the $DP$ system
differs from the nonet symmetric case. At the same time, we will
use $r'_i$, $\bar r'_i$ and $\tilde r'_i$ to predict $D\overline
K$ and $\overline D\,\overline K$ rates, and compare with data
whenever possible.

\subsection{SU(3) Decomposition}

It is instructive to see the phases and angle given in
Eq.~(\ref{eq:solution}) in light of SU(3) decomposition. Let us
consider the $DP$ case first. $D$ is an anti-triplet ($D({\bf
\overline 3})$), while $P$ can be reduced to an octet [$\Pi({\bf
8})$] and a singlet ($\eta_1$). The $D({\bf \overline
3})\otimes\Pi({\bf 8})$ can be reduced into a $\overline{\bold
3}$, a ${\bold 6}$ and a $\overline{\bold 15}$ (see, for example,
\cite{georgi}), while $D({\bf \overline 3})\eta_1$ is another
anti-triplet. Denoting the latter as $\bf \overline 3 {}^\prime$,
it can mix with the $\overline{\bold 3}$ from $D\Pi$ via a
$2\times 2$ symmetric (from time reversal invariance) unitary
matrix ${\cal U}$, which appears already in
Eqs.~(\ref{eq:solution}) and (\ref{eq:U}) . The invariance of the
strong interaction under SU(3) transformation gives
\begin{equation}
 {\cal S}=|{\bf \overline {15}}\rangle \langle {\bf \overline {15}}|
      +e^{2i\delta} |{\bf 6}\rangle\langle {\bf 6}|
      +\left(|{\bf\overline 3}\rangle\,\,\,\,
            |{\bf\overline 3 {}^\prime}\rangle
            \right)
      \cdot{\cal U}
       \cdot \left(
\begin{array}{c}
\langle {\bf \overline 3}|
       \\
\langle {\bf \overline 3{}^\prime}|
\end{array}
\right).
 \label{eq:SSU3}
\end{equation}
It is now clear that, with the choice of vanishing phase in ${\cal
S}_{\bf \overline {15}\,\overline{15}}$, $2\delta$ is the phase of
${\cal S}_{\bf 6\, 6}$ and ${\cal U}$ is the mixing matrix in the
anti-triplet sector. Note that in the master formula
Eq.~(\ref{eq:A=S12A0}) one should use ${\cal S}^{1/2}$. This can
be easily obtained by reducing all phases in the right-hand-side
of the above equation by half.% [see also Eq.~(\ref{eq:U})].

Three remarks are in order.
It is important to emphasize that, by charge conjugation
invariance of the strong interaction, the above $S$-matrix can
also be applied to the $\overline D P$ case with ${\bf
\overline{15}},\,{\bf 6}$ and ${\bf \overline{3} {}^{(\prime)}}$
replaced by ${\bf {15}},\,{\bf \overline6}$ and $\bf
3{}^{(\prime)}$, respectively.
Second, the $D^0\eta_8$ and $D^0\eta_1$ are not physical final
states. The physical $\eta,\,\eta^\prime$ mesons are defined
through
\begin{equation}
\left(
\begin{array}{c}
\eta\\
      \eta^\prime
\end{array}
\right)= \left(
\begin{array}{cc}
\cos\vartheta &-\sin\vartheta\\
\sin\vartheta &\cos\vartheta
\end{array}
\right) \left(
\begin{array}{c}

\eta_8\\
      \eta_1
\end{array}
\right),
\end{equation}
with the mixing angle $\vartheta=-15.4^\circ$
\cite{Feldmann:1998vh}.
In the previous analysis~\cite{Chua:2001br}, $D^0\eta$ was
approximated as $D^0\eta_8$, while $D^0\eta_1$ was ignored. This
corresponds to vanishing mixing angle $\tau = 0$ in ${\cal U}$,
and dropping the ${\bf \overline{3} {}^{\prime}}$.

Before we end this section, let us specify $A^0$ in the above
formulas. We use factorization amplitudes for $A^0$, which
are~\cite{Cheng:2001sc}, for each physical final state~\footnote{
To match the phase convention $T\,|{\rm in}\rangle=|\out\rangle$
(after factoring out the CKM factor) we absorb the factor $i$ of
these factorization amplitudes into the $B$ meson
state~\cite{Cheng:2004ru}. }
 \begin{eqnarray}
 A^f_{D^0\pi^-}&=&V_{cb} V_{ud}^*(T_f+C_f),
 \qquad
 A^f_{D^+\pi^-}=V_{cb} V_{ud}^*(T_f+E_f),
 \nonumber \\
 A^f_{D^0\pi^0}&=&\frac{V_{cb} V_{ud}^*}{\sqrt2}(-C_f+E_f),
 \qquad
 A^f_{D^+_s K^-}=V_{cb} V_{ud}^*~E_f,
 \nonumber \\
 A^f_{D^0\eta_8}&=&\frac{V_{cb} V_{ud}^*}{\sqrt6}(C_f+E_f),
 \qquad
 A^f_{D^0\eta_1}=\frac{V_{cb} V_{ud}^*}{\sqrt3}(C_f+E_f),
 \nonumber\\
 A^f_{D^0 K^-}&=&V_{cb} V_{us}^*(T_f+C_f),
 \qquad
 A^f_{D^+ K^-}=V_{cb} V_{us}^*~T_f,
 \nonumber \\
 A^f_{\overline D {}^0 K^-}&=&V_{ub} V_{cs}^*~(c_f+a_f),
 \qquad
 A^f_{D^- \overline K{}^0}=V_{ub} V_{cd}^*~a_f,
 \nonumber \\
  A^f_{\overline D{}^0\eta_8}&=&\frac{V_{ub} V_{cs}^*}{\sqrt6} (t_f-2 a_f),
 \qquad
 A^f_{\overline D{}^0\eta_1}=\frac{V_{ub} V_{cd}^*}{\sqrt3}(t_f+a_f),
 \nonumber\\
 A^f_{D^-_s\pi^0}&=&\frac{V_{ub} V_{cs}^*}{\sqrt2} t_f,
 \label{eq:Atop}
\end{eqnarray}
where the super- and subscripts $f$ indicate factorization
amplitude, and
\begin{eqnarray}
T_f&=&{G_F\over\sqrt2} \, a^{\rm eff}_1 \, (m_B^2-m_D^2)
           f_P F_0^{BD}(m^2_P),
\nonumber\\
C_f&=&{G_F\over\sqrt2} \, a^{\rm eff}_2 \,
                (m_B^2-m_P^2) f_D F_0^{BP}(m^2_D),
\nonumber\\
E_f&=&{G_F\over\sqrt2} \, a^{\rm eff}_2 \,
                (m_D^2-m_P^2) f_B F_0^{0\to DP}(m^2_B),
\nonumber\\
t_f&=&{G_F\over\sqrt2} \, a^{\rm eff}_1 \, (m_B^2-m_P^2)
           f_D F_0^{BP}(m^2_D),
\nonumber\\
c_f&=&{G_F\over\sqrt2} \, a^{\rm eff}_2 \,
                (m_B^2-m_P^2) f_D F_0^{BP}(m^2_D),
\nonumber\\
a_f&=&{G_F\over\sqrt2} \, a^{\rm eff}_1 \,
                (m_D^2-m_P^2) f_B F_0^{0\to DP}(m^2_B).
\end{eqnarray}
$F_0^{BD(BP)}$ is the $\overline B \to D(P)$ transition form
factor and $F_0^{0\to DP}$ is the vacuum to $DP$ (time-like) form
factor. Some SU(3) breaking effects are included in the decay
constants and various form factors. In subsequent numerical study
we take $E_f=a_f=0$. In this case, it is seen that in the
$\overline B\to D P$ decays, the $a^{\rm eff}_1$ and $a^{\rm
eff}_2$ parameters accompany $F_0^{BD}(0)$ and $F_0^{BP}(m_D^2)$,
respectively. It should be noted that in the quark diagram
approach [or the SU(3) approach] the {\it full} amplitudes (i.e.
without the super- and subscripts $f$) are expressed as in
Eq.~(\ref{eq:Atop}), and one treats $T$ (tree), $C$
(color-suppressed), $E$ (exchange) and $A$ (annihilation) as
complex topological
amplitudes~\cite{Savage:ub,Zeppenfeld:1980ex,Chau:tk,Chau:1990ay,Gronau:1994rj}.

\begin{table}[t!]
\caption{ \label{tab:expt} Summary of experimental results for
$\overline B{}^0 \to DP$
modes~\cite{CLEOBDpi,BaBarBDpi,Belleupdate,Schumann:2005ej,PDG,DsK,D0K0}.}
\begin{ruledtabular}
\begin{tabular}{lcccc}
$\mathcal{B}$ ($\times 10^{-4}$)
      &CLEO
      &BaBar
      &Belle
      &Average\\
\hline $\overline B {}^0\to D^0\pi^0$
        &$2.74^{+0.36}_{-0.32}\pm 0.55$
        &$2.9\pm0.2\pm0.3$
        &$2.31\pm0.12\pm0.23$
        &$2.53\pm0.20$
        \\
$\overline B {}^0\to D^0\eta$
        &--
        &$2.5\pm0.2\pm0.3$
        &$1.83\pm0.15\pm0.27$
        &$2.11\pm0.33$\footnotemark[1]
        \\
$\overline B {}^0\to D^+_s K^-$
        &--
        &$0.32\pm0.12\pm0.08$
        &$0.45^{+0.14}_{-0.12}\pm0.11$
        &$0.38\pm0.13$
        \\
$\overline B {}^0\to D^0\eta^\prime$
        & --
        & $1.7\pm0.4\pm0.2$
        & $1.14\pm0.20^{+0.10}_{-0.13}$
        & $1.26\pm0.23$\footnotemark[2]
        \\
$\overline B {}^0\to D^+ K^-$
        & --
        & --
        & $2.04\pm0.50\pm0.27$
        &$2.0\pm0.6$
        \\
$\overline B {}^0\to D^0\overline K {}^0$
       &--
       &--
       &$0.50^{+0.13}_{-0.12}\pm0.06$
       &$0.50\pm0.14$
\end{tabular}
\footnotetext[1]{The error is scaled by a factor of $S=1.4$.}
\footnotetext[2]{The error is scaled by a factor of $S=1.1$.}
\end{ruledtabular}
\end{table}

\begin{table}[t!]
\caption{ Form factors in covariant light-front models~\cite{LF}.
For $B\to\eta^{\prime}$ form factors the mixing angle and
Clebsch-Gordan coefficients are included.
 \label{tab:formfactor}
}
\begin{ruledtabular}
\begin{tabular}{lclc}
$F_0^{B\pi}(m_{D,D_s}^2)$
  & 0.28
  &$F_0^{BD}(m_{\pi,K}^2)$
  & 0.67
  \\
$F_0^{B\eta}(m_{D,D_s}^2)$
  & 0.15
  &$F_0^{B\eta^\prime}(m_{D,D_S}^2)$
  &0.13 \\
$F^{BK}_0(m^2_{D,D_s})$
  &0.43
  \\
\end{tabular}
\end{ruledtabular}
\end{table}

\section{\label{sec:num} Results and Discussion}

In our numerical study, masses and lifetimes are taken from the
Particle Data Group (PDG)~\cite{PDG}. We use the color suppressed
branching ratios as stated in Table~\ref{tab:expt}. For other
modes, such as $\overline B {}^0\to D^+\pi^-$ and $B^-\to
D^0\pi^-$, $D^0 K^-$ decays, we use PDG values \cite{PDG}. We fix
$V_{ud}= 0.9738$, $V_{us}= 0.2200$, $V_{cb}= 0.0413$,
$V_{cs}=0.996$, $|V_{ub}|=3.67\times 10^{-3}$, and use the decay
constants $f_\pi=$ 131 MeV, $f_{K}=$ 156 MeV~\cite{PDG} and
$f_{D_{(s)}}=$ 200 (230) MeV. Form factors are taken from the
covariant light-front quark model calculation~\cite{LF}, where we
list the relevant values in Table~\ref{tab:formfactor}.

To describe the processes with rescattering from factorization
amplitudes, we have six parameters: the two effective Wilson
coefficients $a^{\rm eff}_1$ and $a^{\rm eff}_2$, and the three
rescattering phases $\delta$, $\theta$ and $\sigma$, and one
mixing angle $\tau$ in ${\cal S}^{1/2}$. In subsection A, these
parameters are fitted with $C = 1$, $S = 0$ $DP$ data, i.e. the
$\overline B \to D^+\pi^-$, $D^0\pi^-$, $D^0\pi^0$, $D^0\eta$,
$D^0\eta'$ and $D^+_s K^-$ rates. We then use the extracted
parameters to predict ($C = 1$, $S = -1$) $\overline B \to D^0
K^-$, $D^+K^-$ and $D^0 \overline K {}^0$ rates and compare with
measurement. Predictions for the ($C = -1$, $S = -1$) $\overline B
\to \overline D\,\overline K$ modes, and the value for $r_B$, are
given in subsection B.

\begin{table}[b!]
\caption{ \label{tab:airi} Fitted parameters of the SU(3) FSI
picture, using $\overline B \to D^+\pi^-$, $D^0\pi^-$, $D^0\pi^0$,
$D^0\eta$, $D^0\eta'$ and $D^+_s K^-$ decay rates (see
Table~\ref{tab:table-br}) as input. There is a two fold ambiguity
(the overall sign of the phases) in the solutions. The SU(3)
phases and mixing are reexpressed in terms of the rescattering
parameters $r_i^\prime$, $\bar r_i^\prime$, $\tilde r_i^\prime$.}
\begin{ruledtabular}
\begin{tabular}{lclc}
      parameter
      & solution
      &parameter
      & solution
      \\
\hline
 $a^{\rm eff}_1$
     &$0.92^{+0.04}_{-0.06}$
     &$a^{\rm eff}_2$
     &$0.22^{+0.12}_{-0.09}$
     \\
 $\delta$
     &$\pm(62.4^{+5.4}_{-5.6})^\circ$
     &$\theta$
     &$\pm(23.8^{+2.0}_{-7.2})^\circ$
     \\
 $\sigma$
     &$\pm(127^{+46.8}_{-77.0})^\circ$
     &$\tau$
     &$(1.7^{+21.3}_{-3.9})^\circ$

     \\
\hline
 $1+i r'_0$
        &$(0.73\pm0.04)\pm (0.44\pm0.02)i$
 &$i r'_e$
        &$(0.27\pm0.04)\mp (0.44\pm0.02)i$
        \\
 $i r'_a$
        &$(0.10\pm0.02)\mp(0.07\pm0.02)i$
 &$i(\bar r'_a+\bar r'_e)$
        &$(-0.05^{+0.11}_{-0.13})\pm (0.01^{+0.16}_{-0.11})i$
        \\
 $1+i \tilde r'_0+i\frac{\tilde r'_e+\tilde r'_a}{3}$
        &$(-0.61^{+1.29}_{-0.39})\pm(0.79^{+0.21}_{-1.37})i$
 &$i(r'_a+r'_e)$
        &$(0.37\pm0.06)\mp (0.51\pm0.03)i$
        \\
 $1+i r'_0+i\frac{r'_e+ r'_a}{3}$
        &$(0.86\pm0.02)\pm(0.27^{+0.01}_{-0.02})i$
\end{tabular}
\end{ruledtabular}
\end{table}

\subsection{FSI Effects on $\overline B\to DP$ and $D\overline K$ Rates}

Taking the $\overline B \to D^+\pi^-$, $D^0\pi^-$, $D^0\pi^0$,
$D^0\eta$, $D^0\eta'$ and $D^+_s K^-$ rates as input, we fit for
$a^{\rm eff}_{1,2}$ and the FSI phases and mixing. The fitted
parameters are given in Table~\ref{tab:airi}, where errors are
propagated from the experimental errors. The factorization rates
and the predicted $D\overline K$ rates are compared to
experimental results in Table~\ref{tab:table-br}. The
factorization rates are determined by using the $a^{\rm
eff}_{1,2}$ from the fit, but setting all FSI phases and mixing
angle to zero. Unitarity is then implied automatically, i.e. sum
of rates within coupled modes are unchanged by FSI.

Table~\ref{tab:table-br} illustrates the effect of FSI. $\overline
B{}^0\to D^0 h^0$ (where $h^0 = \pi^0$, $\eta,\ \eta^\prime$)
rates are fed mostly by $\overline B{}^0\to D^+\pi^-$. Since these
rescattering parameters are extracted from CP even measurements,
the overall sign of phases is undetermined.

The FSI contributions for the $D_s^+ K^-$, $D^0\pi^0$ and
$D^0\eta_8$ rates from $D^+\pi^-$ rescattering are governed by
$r'_a$, $r'_a-r'_e$ and $r'_a+r'_e$, respectively. The strength of
${\cal B}(\overline B{}^0\to D_s^+K^-) \simeq 4\times 10^{-5}$
implies that $r^{\prime}_a$ cannot be too small, i.e. $|
r'_a|\simeq \sqrt{{\mathcal B}(D_s^+ K^-)/{\mathcal
B}(D^+\pi^-)}\simeq 0.12$. On the other hand the FSI enhances
$D^0\pi^0$ rate from $0.5\times 10^{-4}$ to $2.5\times 10^{-4}$.
Comparing these two modes, we have $|r'_e-r'_a|>|r'_a|$.
Consequently, through the analysis of the two above modes, the
size of FSI contribution to $D^0\eta_8$ is roughly determined. In
fact, the FSI contribution alone already gives ${\mathcal
B}(D^0\eta_8)\simeq 2\times 10^{-4}$, and after interference with
the short distance contribution, one gets ${\mathcal
B}(D^0\eta_8)\simeq 3\times 10^{-4}$. To fit the $D^0\eta$ data,
the $D^0\eta_8$ amplitude interferes destructively, through
$\sigma>90^\circ$, with the $D^0\eta_1$ amplitude, while at the
same time the $D^0\eta'$ amplitude gets enhanced through
constructive interference. Although the $D^0\eta_1$ amplitude is
small and does not be enhanced in FSI (due to the smallness of
$\bar r'_{e,a}$), it still affects $D^0\eta$ and $D^0\eta'$ rates
through interferences.

\begin{table}[t!]
\caption{ \label{tab:table-br} The branching ratios of various
$\overline B\to DP$ and $D\overline K$ modes in $10^{-4}$ units.
The second and third columns compare experiment with factorization
model, and the last column gives the FSI results. The
factorization results are obtained by using the same set of
parameters but with FSI phases set to zero.}
\begin{ruledtabular}
\begin{tabular}{lccc}
 Mode
      &$\mathcal{B}^{\rm exp}$ ($10^{-4}$)
      &$\mathcal{B}^{\rm fac}$ ($10^{-4}$)
      &$\mathcal{B}^{\rm FSI}$ ($10^{-4}$)
      \\
\hline
 $D^0\pi^-$
        & $49.8\pm2.9$
        & input %$49.8\pm2.9$
        & input %$(49.8\pm2.9)$
        \\
\hline $D^+\pi^-$
        & $27.6\pm2.5$
        & $33.0^{+3.0}_{-4.3}$
        & input %(27.6\pm2.5)$
        \\
$D^0\pi^0$
        & $2.53\pm 0.20$
        & $0.51^{+0.72}_{-0.34}$
        & input %$(2.53\pm0.20)$
        \\
$D^+_s K^-$
        & $0.38\pm0.13$
        & 0
        & input %$(0.38\pm0.13)$
        \\
$D^0\eta$
        & $2.11\pm0.33$
        & $0.29^{+0.41}_{-0.20}$
        & input %$(2.11\pm0.33)$
        \\
$D^0\eta'$
        & $ 1.26\pm 0.26 $
        & $0.18^{+0.26}_{-0.12}$
        & input %$(1.26\pm0.26)$
        \\
\hline $D^0 K^-$
        & $3.7\pm0.6$
        & $3.91^{+0.37}_{-0.32}$
        & $3.91^{+0.37}_{-0.32}$
        \\
\hline
 $D^+K^-$
        & $2.0\pm 0.6$
        & $2.38^{+0.21}_{-0.31}$
        & $1.78^{+0.20}_{-0.17}$
        \\
$D^0 \overline K{}^0$
        & $0.50\pm0.14$
        & $0.12^{+0.17}_{-0.08}$
        & $0.73^{+0.08}_{-0.10}$
        \\
\end{tabular}
\end{ruledtabular}
\end{table}

Utilizing the SU(3) framework, we can predict the results for $C=
1$, $S = -1$ $D\overline K$ modes. The prediction for $\overline B
\to D^0 K^-$, $D^+K^-$ and $D^0\overline K {}^0$ rates are given
in Table~\ref{tab:table-br}, where the experimental data are also
listed. These rates were not used in the fit for $a_i$ and FSI
parameters. We see that the $D^0 K^-$, $D^+K^-$ rates are in good
agreement with data. The color-suppressed $D^0\overline K {}^0$
rate is a bit larger than data, but still in reasonable agreement.

From Table~\ref{tab:airi} we observe that $1+i
r'_0\simeq0.85~e^{\pm i31^\circ}$ is almost perpendicular to $i
r'_e\simeq 0.52~e^{\mp i58^\circ}$. We do not know the reason for
this orthogonality, but this implies that the FSI amplitude from
$D^+ K^-\to D^0 \overline K {}^0$ rescattering is almost
perpendicular to the one from $D^0 \overline K {}^0 \to D^0
\overline K {}^0$ rescattering [c.f. Eq.~(\ref{eq:A=S12ADK})].
Consequently, we have
 \begin{equation}
 {\mathcal B}(D^0\overline K {}^0)
 \simeq |1+i r'_0|^2 {\cal B}^{\rm fac}(D^0 \overline K {}^0)+|ir'_e|^2{\cal B}^{\rm fac}(D^+ K^-),
 \label{eq:sumDK}
 \end{equation}
which gives a very good approximation of the result shown in
Table~\ref{tab:table-br}. As we shall see, a relation similar to
Eq.~(\ref{eq:sumDK}) holds for the $B^-\to \overline D {}^0 K^-$
case. From Eq.~(\ref{eq:sumDK}) we see that a $\sim 15\%$
reduction of $a^{\rm eff}_1 r'_e$ from its central value can
reproduce the current ${\mathcal B}(D^0\overline K {}^0)$ central
value. In fact, a smaller ${\mathcal B}(D^0\overline K {}^0)\simeq
0.5\times 10^{-4}$ was predicted~\cite{Chua:2001br} with a smaller
$\delta$ extracted~\footnote{
 In place of $\delta$,
a different notation $\delta'$ was used in \cite{Chua:2001br}.
 }
from earlier $D^0 h^0$ data, so the measurements probably have yet
to settle. Note that the rate of the color-allowed $D^0 K^-$ mode
is not affected by the quasi-elastic rescattering, just like
$D^0\pi^-$.

\begin{table}[b!]
\caption{ \label{tab:rBdBexpt} Summary of experimental results for
$r_B$, $\delta_B$ and $\phi_3$ %$\gamma$
in the $DK$ Dalitz method~\cite{HFAG,gammaDK}. For the phase
convention adopted see footnote~\ref{fn:phase}.}
\begin{ruledtabular}
\begin{tabular}{lcc}
      &Belle
      &BaBar
      \\
\hline $r_B$
        &$0.21\pm0.08\pm0.03\pm0.04$
        &$< 0.19$ (90\% CL)
        \\
$\delta_B$
        &$-23^\circ\pm 19^\circ\pm11^\circ\pm21^\circ$
        &$-66^\circ\pm 41^\circ\pm 8^\circ \pm 10^\circ$
        \\
$\phi_3 (\gamma)$
        &$64^\circ\pm 19^\circ\pm  13^\circ\pm  11^\circ$
        &$70^\circ\pm 44^\circ\pm  10^\circ\pm 10^\circ $
\end{tabular}
\end{ruledtabular}
\end{table}

\begin{table}[t!]
\caption{ \label{tab:table-br1} Predictions for $B^-\to\overline
D\, \overline K$ rates. Experimental limits~\cite{PDG} are shown
in the second column. The third and fourth columns are
factorization and FSI results, respectively, using the same
parameters as Table~\ref{tab:airi}. }
\begin{ruledtabular}
\begin{tabular}{lccc}
 Mode
      &$\mathcal{B}^{\rm exp}$ ($10^{-5}$)
      &$\mathcal{B}^{\rm fac}$ ($10^{-5}$)
      &$\mathcal{B}^{\rm FSI}$ ($10^{-5}$)
      \\
\hline
 $\overline D {}^0 K^-$
        & --
        &$0.17^{+0.23}_{-0.11}$
        &$0.28^{+0.23}_{-0.15}$
        \\
 $D^- \overline K{}^0$
        & --
        & 0
        &$0.05^{+0.06}_{-0.03}$
        \\
$D_s^-\pi^0$
        & $<20$
        & $0.77^{+0.07}_{-0.10}$
        & $0.59^{+0.06}_{-0.05}$
        \\
$D_s^-\eta$
        & $<50$
        & $0.46^{+0.04}_{-0.06}$
        & $0.17^{+0.30}_{-0.09}$
        \\
$D_s^-\eta'$
        & --
        & $0.30\pm0.03$
        & $0.58^{+0.12}_{-0.26}$
        \\
\end{tabular}
\end{ruledtabular}
\end{table}

\subsection{$\overline B\to \overline D\,\overline K$ Rates
 and Prediction of $r_B$}

In this subsection, the $\overline B\to\overline D\,\overline K$
rates and $r_B$, $\delta_B$ are predicted and compared with data.

Table~\ref{tab:rBdBexpt} gives the current experimental results on
$r_B$, $\delta_B$ and $\phi_3/\gamma$ from~\footnote{
 Note that we use the
phase convention $CP\,|D^0\rangle=-|\overline D {}^0\rangle$ and,
consequently, our $\delta_B$ is related to those in
\cite{HFAG,gammaDK} by a $\delta_B-\pi$
transformation\label{fn:phase}.
 }
the $DK$ Dalitz method~\cite{HFAG,gammaDK}. Our predictions for
$\overline B\to \overline D{}^0 K^-$, $D^- \overline K{}^0$,
$D_s^-\pi^0$, $D_s^-\eta$ and $D_s^-\eta'$ decay rates are given
in Table~\ref{tab:table-br1}. The sum of rates is unchanged in the
presence of FSI. Note that the first two modes have $I=0,1$
components, the third mode is purely $I=1$ while the last two
modes are purely $I=0$. Rescattering between $D_s^-\pi^0$ and
$D_s^-\eta^{(\prime)}$ is forbidden by isospin. Consequently, the
$D_s^-\eta$ and $D_s^-\eta^\prime$ rates do not receive any
contribution form the $D_s^-\pi^0$ mode. These two modes also do
not rescatter much from $D^- \overline K{}^0$ and $\overline D
{}^0 K^-$, as the rescattering parameters are either suppressed by
Clebsch-Gordan coefficients, or by the smallness of $\bar
r'_e+\bar r'_a$ (c.f. Table~\ref{tab:airi}). Thus, these two modes
basically rescatter among themselves (as one may check that the
sum of their rates are roughly conserved under FSI).

The $D^-_s\eta$ and $D^-_s\eta'$ rates are reduced and enhanced,
respectively, through FSI between themselves. This is due to the
destructive and constructive interference effects of $D^-_s\eta_8$
and $D^-_s\eta_1$ in the $\sigma>90^\circ$ case as required from
the $D^0\eta^{(\prime)}$ data. Note that within error we can also
have $\sigma<90^\circ$, so ${\mathcal B}(D^-_s\eta)>{\mathcal
B}(D^-_s\eta')$ is not ruled out.

For the first three modes, the dominant source of rescattering is
$D_s^-\pi^0$. It feeds $\overline D {}^0 K^-$ and $D^- K^0$
through $r'_e/\sqrt2$ and $-r'_e/\sqrt2$, respectively. Note that
these ${\cal S}^{1/2}$ matrix elements are similar to the
$D^+K^-\to D^0 \overline K {}^0$ rescattering matrix element
except for the $1/\sqrt2$ factor. Furthermore, we see from
Table~\ref{tab:airi} that $1+i r'_0+i r'_a\simeq 0.97 e^{\pm
i32^\circ}$ is perpendicular to $i r'_e\simeq 0.52~e^{\mp
i58^\circ}$. Consequently, the FSI amplitude from $D^-_s\pi^0\to
\overline D {}^0 K^-$ rescattering is orthogonal to the one from
$\overline D {}^0 K^-\to \overline D {}^0 K^-$ rescattering [c.f.
Eq.~(\ref{eq:FSIDbarK})], and the rate of $\overline D {}^0 K^-$
is roughly given by
 \begin{equation}
 {\mathcal B}(\overline D {}^0 K^-)
 \simeq {\mathcal B}^{\rm fac}(\overline D {}^0 K^-)+ \left|\frac{ir'_e}{\sqrt2}\right|^2{\mathcal B}^{\rm fac}(D^-_s \pi^0)
 \simeq {\mathcal B}^{\rm fac}(\overline D {}^0 K^-)+0.13~{\mathcal B}^{\rm fac}(D^-_s \pi^0),
 \label{eq:sumDbarK}
 \end{equation}
which is analogous to Eq.~(\ref{eq:sumDK}) for the $D^0\overline
K{}^0$ case. Since the $D^0 \overline K {}^0$ and $\overline D
{}^0 K^-$ modes have similar relations and the FSI contributions
are governed by the same parameter $ir'_e$, the $D^0\overline
K{}^0$ mode may provide some estimation of the FSI effect in the
$\overline D {}^0 K^-$ case. For example, if the current central
value of $5\times 10^{-5}$ for $D^0 \overline K {}^0$ holds, we
would need roughly a 15\% reduction in $a_1^{\rm eff} r'_e$ as
discussed in the previous section. This corresponds to a 10\%
reduction in the predicted $\overline D {}^0 K^-$ rate or,
equivalently, a 5\% reduction in $|A_{\overline D {}^0 K^-}|$.
Such a variation is within the error shown in
Table~\ref{tab:table-br1}.

The $B^-\to D^- \overline K{}^0$ decay is a pure annihilation
decay mode. Its decay rate can give us an idea of the size of the
annihilation amplitude. We set the short distance annihilation
contribution to zero. Its rate then comes mainly from $D^-_s
\pi^0$ rescattering, which interferes destructively with the
rescattering contribution from $\overline D {}^0 K^-$. In terms of
topological amplitudes we have $(a/t)_{\overline D\, \overline
K}=0.21\pm0.08$ and $\arg\,(a/t)_{\overline D\, \overline
K}=\pm(104.8^{+14.3}_{-10.2})^\circ$, where $a$ and $t$ are the
full annihilation and tree amplitudes, respectively [see
Eq.~(\ref{eq:Atop}) and subsequent discussion]. The
$(a/t)_{\overline D\, \overline K}$ ratio is numerically close to
$a^{\rm eff}_2/a^{\rm eff}_1$. The long distance annihilation
amplitude cannot be neglected.

The $A(B^- \to \overline D {}^0 K^-)$ and $A(B^- \to D^0 K^-)$
amplitude ratio gives $r_B$ and $\delta_B$, and the values are
shown in Table~\ref{tab:rB} and compared with experiment. The
$r_B$ parameter [c.f. Eq.~(\ref{eq:rB})] governs the strength of
the interference effect that is essential for $\phi_3/\gamma$
determination in the GLW, ADS and $DK$ Dalitz methods. We see that
our $r_B=0.09\pm0.02$ prefers the BaBar result over the Belle
result, while $\delta_B$ is in agreement with both BaBar and Belle
results. Our $r_B$ value is also in good agreement with the fit
from UT$_{fit}$ group, obtained by using all three methods of GLW,
ADS and $DK$ Dalitz analysis. The smallness of $r_B$ implies that
we need more $B$ data to determine $\phi_3$.

\begin{table}[t!]
\caption{ \label{tab:rB} Factorization and FSI results on $r_B$,
$\delta_B$ with $|V_{ub}|=3.67\times10^{-3}$, and compared to the
experimental results~\cite{gammaDK,UTfit,HFAG}. For the phase
convention adopted see footnote~\ref{fn:phase}.}
\begin{ruledtabular}
\begin{tabular}{lccc}
      &Expt
      &fac
      &FSI
      \\
\hline
 $r_B$
        &$0.21\pm0.08\pm0.03\pm0.04$ (Belle)
               &$0.07\pm0.03$
        &$0.09\pm0.02$
        \\
        &$< 0.19$ (90\% CL) (BaBar)
        \\
        &$0.10\pm0.04$ (UT$_{fit}$)
        \\
 $\delta_B$
        & $-23^\circ\pm 19^\circ\pm11^\circ\pm21^\circ$ (Belle)
        &$0$
        &$\mp(19.9^{+25.1}_{-13.9})^\circ$
        \\
        & $-66^\circ\pm 41^\circ\pm 8^\circ \pm
          10^\circ$ (BaBar)
        &
        &
\end{tabular}
\end{ruledtabular}
\end{table}

It is interesting to note that the ratios $\sqrt{{\mathcal
B}(D^0\pi^0)/{\mathcal B}(D^+\pi^-)}$, $\sqrt{{\mathcal B}(D^0
\overline K {}^0)/{\mathcal B}(D^+ K^-)}$ and $\sqrt{{\mathcal
B}(\overline D {}^0 K^-)/{\mathcal B}(D^-_s \pi^0)}$ are enhanced
by roughly $2.44$, $2.85$ and $1.47$, respectively, from their
factorization values. These enhancements are sometimes interpreted
as~\cite{Neubert:2001sj,Cheng:2001sc} enhancement in $|a_2/a_1|$.
However, the measured ratios are nonuniversal. In particular, we
would have a larger $r_B$ if we straightforwardly apply the
$|a_2/a_1|$ ratio from the $D\pi$ and $DK$
systems~\cite{Gronau:2002mu}. So let us try to understand the
smallness of $r_B$ or, equivalently, the smallness of the
enhancement in $\sqrt{{\mathcal B}(\overline D {}^0 K^-)/{\mathcal
B}(D^-_s \pi^0)}$.

In the factorization approach we have
 \begin{equation}
 r^{\rm fac}_B\simeq
 \left|\frac{V_{ub}V_{cs}}{V_{cb}V_{us}}\right|
  \frac{a^{\rm eff}_2}{a^{\rm eff}_1 X_{DK}+a^{\rm eff}_2}
  \simeq 0.07
 \end{equation}
   with
$X_{DK}\equiv (m_B^2-m_D^2)f_K F_0^{BD}(m_K^2)/[(m_B^2-m_K^2)f_D
F_0^{BK}(m_D^2)]\simeq 1.08$. This $r^{\rm fac}_B$ agrees with the
common estimation. In Table~\ref{tab:rB} we see that $r_B$ is
indeed enhanced from its factorization value but the enhancement
is mild. Although the $\overline D {}^0 K^-$ rate is enhanced by
70\% from its factorization value, the amplitude is only enhanced
by 30\%. The rate of the color-allowed $D^0 K^-$ mode is not
affected by the quasi-elastic rescattering. Consequently, $r_B$
does not differ much from its factorization prediction. Note that
although the predicted $\overline D {}^0 K^-$ rate has a large
error, the error in $r_B$ is much reduced as it is a ratio and,
furthermore, a ratio of amplitudes.

It is instructive to compare $B^- \to \overline D{}^0 K^-$ with
the color suppressed modes $\overline B{}^0\to D^0\pi^0$ and
$D^0\overline K^0$. Before rescattering, i.e. at factorization
level, we have
 \begin{eqnarray}
  \frac{A^f_{D^0\pi^0}}{A^f_{D^+\pi^-}} &\simeq&
  \frac{a^{\rm eff}_2}{\sqrt2 a^{\rm eff}_1 X_{D\pi}} \simeq 0.13,
  \nonumber\\
  \frac{A^f_{D^0\overline K {}^0}}{A^f_{D^+ K^-}} &\simeq&
  \frac{a^{\rm eff}_2}{a^{\rm eff}_1 X_{DK}} \simeq 0.23,
  \nonumber\\
  \frac{A^f_{\overline D^0 K^-}}{A^f_{D^-_s\pi^0}} &\simeq& \sqrt2~Y_{DK}
  \frac{a^{\rm eff}_2}{a^{\rm eff}_1}\simeq 0.46
 \end{eqnarray}
with $X_{DP}\equiv (m_B^2-m_D^2) f_P
F_0^{BD}(m_P^2)/[(m_B^2-m_P^2) f_D F_0^{BP}(m_D^2)]$, which gives
$X_{D\pi}\simeq 1.38$, and $Y_{DK}\equiv f_D
F_0^{BK}(m_D^2)/[f_{D_s} F_0^{B\pi}(m_{D_s}^2)]\simeq 1.33$. The
rates of the sources $D^+\pi^-$ and $D^+ K^-$ are much larger than
the corresponding color suppressed modes, hence the effects of FSI
are prominent. For $D^+\pi^-$, it can be traced to the source mode
being relatively enhanced by $X_{D\pi}$, while the color
suppressed $D^0\pi^0$ is further suppressed by $1/\sqrt 2$ in the
$\pi^0$ wavefunction. In the case of $B^-\to \overline D {}^0
K^-$, however, the situation is reversed. The color-suppressed
mode $\overline D{}^0 K^-$ is relatively enhanced by $Y_{DK}$,
while the $D^-_s\pi^0$ source mode receives the $1/\sqrt 2$
wavefunction suppression. Thus, the factorization rate for
$\overline D{}^0 K^-$ differs less from the $D_s^- \pi^0$ source
rate, and the enhancement of $r_B$ through FSI is mild.
%, while the rescattering parameter $r'_e$ is not necessary
% suppressed [see Eq.~(\ref{eq:sumDbarK}) and Table~\ref{tab:airi}].
Since the FSI effect is not simply multiplicative [see
Eqs.~(\ref{eq:sumDK}) and (\ref{eq:sumDbarK})], we do not have a
simplistic universal enhancement in $|a_2/a_1|$.

\subsection{Discussion}

Let us first offer some remarks on the fitted results of
Tables~\ref{tab:airi} and \ref{tab:table-br}. The effective Wilson
coefficients $a^{\rm eff}_1=0.92^{+0.04}_{-0.06}~
[F^{BD}_0(0)/0.67]$ and $a^{\rm eff}_2=0.22^{+0.12}_{-0.09}~
[F^{B\pi}_0(m_D^2)/0.28]$ obtained in our fit to $DP$ data agree
well with $|a_2|=0.26\pm0.02$ from fit to $B\to J/\psi K$
data~\cite{Cheng:1999kd}, and with the range of $a^{\rm
eff}_1\simeq 1$, $a^{\rm eff}_2\simeq 0.2$--0.3 from various modes
\cite{Neubert:1997uc,Cheng:1999kd}. The ratio $a^{\rm
eff}_2/a^{\rm eff}_1\simeq 0.24$ is close to the one used in our
previous analysis \cite{Chua:2001br}.

We find from data that a small mixing angle $|\tau|\ll 1$ is
preferred. Thus, the approximation of treating $\eta$ as $\eta_8$
and ignoring $\eta_1$ taken in our previous
analysis~\cite{Chua:2001br} is basically valid, and the FSI phases
$\delta$ and $\theta$ correspond to those in \cite{Chua:2001br}.
In fact, the values for the phases $\delta\simeq\pm62^\circ$ and
$\theta\sim24^\circ$ are consistent with the previous results of
$\pm 48^\circ$ and $\pm 25^\circ$, respectively. We note that, as
a consequence of the smallness of $\tau$, the phase $\sigma$ is
less constrained from the present $D^0\eta^{(\prime)}$ data.

From the $r'_i$ values given in Table~\ref{tab:airi}, we see that
exchange rescattering is dominant over annihilation rescattering.
Comparing $\bar r'_i$ and $\tilde r'_i$ with $r'_i$, which should
be identical in the U(3) limit, we see that U(3) is not a very
useful limit for these modes.

In the present work the overall sign of FSI phases cannot be
determined. We may obtain some information comparing to other
work. For example, a pole model calculation~\cite{Cheng:2004ru}
with some inputs, such as the Wilson coefficients $a^{\rm
eff}_{1,2}$ and cut-offs for form factors in strong interaction
vertices, or the pQCD approach can give $D^0h^0$ rates in good
agreement with $D^0 h^0$ data and our results.
For the $D\pi$ system, we have
 \begin{equation}
 \left(\frac{C-E}{T+C}\right)_{D\pi}= \frac{-\sqrt2 A_{D^0\pi^0}}{A_{D^0\pi^-}}\simeq 0.33 e^{\mp 91^\circ i},
 \quad
 \left(\frac{C-E}{T+E}\right)_{D\pi}=\frac{-\sqrt2 A_{D^0\pi^0}}{A_{D^+\pi^-}}\simeq 0.43 e^{\mp
 116^\circ i},
 \label{eq:TCE}
 \end{equation}
where negative (positive) phase corresponds to the case of
positive (negative) $\delta$, $\sigma$ and $\theta$. Comparing to
$(C-E)/(T+C)=0.33~e^{-50^\circ i}$~\cite{Cheng:2004ru}, $0.34
e^{-92^\circ i}$~\cite{pQCD} and $(C-E)/(T+E)=0.40~e^{-67^\circ
i}$~\cite{Cheng:2004ru}, $0.32~e^{-111^\circ i}$~\cite{pQCD}, the
case of positive $\delta$, $\sigma$ and $\theta$ phases is
preferred. It is interesting to note that, since the amplitudes
for color suppressed modes in the $D\pi$ system are fed dominantly
from the same amplitude $A_{D^+\pi^-}$, by neglecting contribution
from $a_2$, we have
 \begin{equation}
 \left(\frac{C-E}{T+C}\right)_{D\pi}\simeq \frac{ir'_e-i r'_a}{1+ir'_0+i r'_e}\simeq 0.41 e^{\mp 65^\circ i},
 \quad
 \left(\frac{C-E}{T+E}\right)_{D\pi}\simeq\frac{i r'_a}{1+i r'_0+ir'_a}\simeq 0.49 e^{\mp
 89^\circ i},
 \end{equation}
which give estimations within 30\% errors. %}

In the extraction of $\phi_3$ from $\overline B\to D\overline K$,
$\overline D\,\overline K$ decays the sign of the strong phase
$\delta_B$ can be determined because one is making a $CP$
violation study. The above preferable case of positive FSI phases
$\delta$, $\sigma$ and $\theta$ leads to a negative $\delta_B$
(c.f. Table~\ref{tab:rB}), which is supported by the measured
central values.

Let us turn to the question of $r_B$ determination, and compare
with other approaches.

We fitted six parameters, two effective Wilson coefficients, three
FSI phases and one mixing angle, from rates of six modes,
$D^0\pi^-$, $D^+\pi^-$, $D^0\pi^0$, $D^+_s K^-$, $D^0\eta$ and
$D^0\eta'$. These parameters are fully determined as the number of
unknowns equals that of input and, consequently, the errors in
parameters are propagated from data errors. To keep the above
features we do not include more modes, such as the three
$D\overline K$ modes as input. In principle, we can also include
them in the fit. However, as discussed already after
Eq.~(\ref{eq:sumDbarK}), the $r_B$ obtained in the new fit should
be consistent with the one given here within errors.

One can extract the topological amplitudes from the $D^0 h^0$ data
as in Ref.~\cite{Chiang:2002tv}. However, it is not clear how to
apply these amplitudes to the $\overline D P$ system as the two
topologies are not identical; for example, the form factor
dependence in $t_f$ and $T_f$ in Eq.~(\ref{eq:Atop}) are
different. Furthermore, the annihilation amplitude, which turns
out to be non-negligible in the $\overline D P$ system, cannot be
extracted in the $D^0 h^0$ system.

In pQCD approach the $D^0 h^0$ rates are explained through the
enhancement in $C$ from the incomplete cancellation in the
non-factorization contribution~\cite{pQCD}. Similar mechanism may
lead to an enhanced $r_B$. Recently, a calculation in the pQCD
approach gives $r_B=0.093$~\cite{Keum}, which is close to our
result.

Color allowed $\overline B \to D\pi$ modes can be calculated in
the framework of QCD factorization~\cite{Beneke:2000ry}, but the
color suppressed decay amplitudes with the emission of a heavy
meson cannot. A process dependent $a_2$ approach is
used~\cite{Cheng:2001sc,Neubert:2001sj}. For a comparison to the
present approach, see Ref.~\cite{Chua:2001br}.
In this vein, soft colinear effective theory (SCET) have received
some attention lately. Although SCET cannot predict the $D^0 h^0$
rate, the prediction on the similarity of $D^{*0} h^0$ and
$D^0h^0$ phase and the ${\mathcal B}(D^0\eta')/{\mathcal
B}(D^0\eta)$ ratio agree with data within error~\cite{SCET}. It
would be interesting to see the SCET prediction on $r_B$.
We note that the strong phases extracted form $D^{*0}h^0$ in our
FSI approach are similar to those in $D^0 h^0$ \cite{Chua:2001br},
in agreement with SCET. Consequently, $r_B$ in the $\overline
D{}^* K^-$ system could be similar to that presented here. Indeed,
$r_B=0.09\pm0.04$~\cite{UTfit} in the $\overline D{}^* K^-$ system
is given by the UT$_{fit}$ group and is close to our estimation
for $\overline B \to \overline DK^-$.

\section{Conclusion}

We study quasi-elastic rescattering effects in $\overline B\to
DP$, $D\overline K$ and $\overline D\, \overline K$ modes. The
updated $\overline B{}^0 \to D^0\pi^0$, $D_s^+K^-$, $D^0\eta$,
$D^0\eta'$ data, together with $D^+\pi^-$ and $D^0\pi^-$, are used
to extract $a_{1,2}^{\rm eff}$ and four rescattering parameters.
We find the effective Wilson coefficients $a^{\rm eff}_1\simeq
0.92$, $a^{\rm eff}_2\simeq 0.22$, the strong phases $\delta\simeq
62^\circ$, $\theta=24^\circ$, $\sigma\simeq 127^\circ$ and mixing
angle $\tau\simeq 2^\circ$. The values of $\delta$ and $\theta$
are close to our previous results~\cite{Chua:2001br} ignoring
$D\eta_1$. The smallness of $\tau$ implies small mixing of
$D^0\eta_1$ with other $DP$ modes, hence our previous
approximation is valid. The predicted $B^-\to D^0K^-$ and
$\overline B{}^0 \to D^+ K^-$, $D^0\overline K {}^0$ rates are in
agreement with data. The formalism can be applied to $\overline
B\to \overline D\, \overline K$ modes, and the rates for
$\overline D {}^0 K^-$, $D^- \overline K{}^0$, $D_s^-\pi^0$,
$D_s^-\eta$ and $D_s^-\eta'$ modes are predicted. In particular,
we predict $r_B=0.09\pm0.02$, which agrees with the UT$_{fit}$
extraction~\cite{UTfit} and a recent pQCD result~\cite{Keum}. Our
$r_B$ value prefers the lower value of the BaBar experiment and
disfavors the Belle result, extracted from the $\phi_3/\gamma$ fit
to $B^-\to \{D^0,\, \overline D{}^0\}K^-$ data using the $DK$
Dalitz method.

\begin{acknowledgments}
 We would like to thank Hsiang-nan Li for useful discussion.
This work is supported in part by the National Science Council of
R.O.C. under Grants NSC-93-2811-M-001-059, NSC-93-2112-M-001-053
and NSC-93-2112-M-002-020.
\end{acknowledgments}

\end{document}